\documentclass[amsmath,amssymb,nofootinbib,onecolumn,prd,11pt]{revtex4}
\def\e{\mathrm{e}}
\def\p{\partial}
\def\cL{\mac{L}}
\def\G{\mac{G}}
\renewcommand{\a}{\alpha}
\renewcommand{\th}{\theta}
\newcommand{\tb}{\bar\theta}
\newcommand{\s}{\sigma}
\newcommand{\diffp}[1]{\frac{{\partial}}{{\partial}#1}}
\newcommand{\be}{\begin{equation}}
\newcommand{\ee}{\end{equation}}
\newcommand{\ra}{\rightarrow}
\renewcommand{\i}{\mathrm i}
\def\nn{\notag}
\def\d{\mathrm d}
\def\cD{\mac{D}}
\def\bcD{\bar{\mac{D}}}
\def\Phid{\Phi^\dagger}
\def\zf{{\text{0f}}}
\def\Db{\bar{D}}
\newcommand{\mac}[1]{\mathcal{#1}}

\begin{document}

\begin{titlepage}

\title{Supersymmetric $k$-defects}

\author{Michael Koehn}
\email[]{koehn@physics.upenn.edu}

\author{Mark Trodden}
\email[]{trodden@physics.upenn.edu}
\affiliation{Center for Particle Cosmology, Department of Physics \& Astronomy,\\University of Pennsylvania, 209 South 33rd St, Philadelphia, PA 19104, U.S.A.}

\begin{abstract}
In supersymmetric theories, topological defects can have nontrivial behaviors determined purely by whether or not supersymmetry is restored in the defect core. A well-known example of this is that some supersymmetric cosmic strings are automatically superconducting, leading to important cosmological effects and constraints. We investigate the impact of nontrivial kinetic interactions, present in a number of particle physics models of interest in cosmology, on the relationship between supersymmetry and supercurrents on strings. We find that in some cases it is possible for superconductivity to be disrupted by the extra interactions.

\end{abstract}
\maketitle

\end{titlepage}

\section{Introduction}
Topological defects can arise in any spontaneously broken theory in which the vacuum manifold - the space of vacua of the theory - is topologically nontrivial. Classic examples are magnetic monopoles, cosmic strings and domain walls, with all of these being realized by nature in laboratory systems. In a cosmological setting, in which spontaneously broken symmetries are restored at high temperatures in the early universe, defects can form during the cooling of the cosmos, with implications for its evolution and for other cosmological observables (see, for example~\cite{Vilenkin:2000jqa}).

A crucial fact determining how a network of defects, particularly cosmic strings, evolves is whether the strings themselves carry supercurrents or not. Superconducting cosmic strings have alternative ways to lose energy beyond purely gravitational ones, and can form quasi-stable remnants among other unusual properties~\cite{Witten:1984eb}. While the question of whether strings are superconducting is often decided by the particle content and couplings one chooses to include in the theory, there is, interestingly, a popular class of theories for which supercurrents arise naturally. In supersymmetric theories, cosmic strings for which supersymmetry is restored in the core frequently must carry supercurrents as a consequence of their supersymmetric nature~\cite{Davis:1997bs,Davis:1997ny}. This allows us to place significant constraints on the symmetry structure of theories at a variety of energy scales~\cite{Brandenberger:1996zp}.

In recent years, theorists have become fascinated by a set of non-minimal derivative interactions in field theories as possible ways to address a number of outstanding questions posed by cosmology, specifically in constructing models of the early universe, and in attempting to explain late-time cosmic acceleration~\cite{Joyce:2014kja}. The simplest examples of these non-minimal interactions are given by the so-called $k$-essence or $k$-inflation models~\cite{ArmendarizPicon:2000ah,ArmendarizPicon:1999rj}, and even more exotic examples are provided by the galileon-type interactions~\cite{Nicolis:2008in} that one finds in some extra-dimensional theories~\cite{Dvali:2000hr} and in massive gravity~\cite{deRham:2010ik,deRham:2010kj}.  A natural question to ask, therefore, is whether these interactions affect the robust connection between supersymmetry and the superconductivity of topological defects.

In this letter we answer this question for a class of supersymmetric theories generalizing the $N=1$ models studied in~\cite{Davis:1997bs}. Defects in non-supersymmetric $P(X)$ theories (where $X$ is defined as the canonical kinetic term) - $k$-defects - have been studied previously~\cite{Babichev:2006cy,Babichev:2007tn,Andrews:2010eh}. 
Here, we introduce a supersymmetric extension of this gauge-invariant higher-derivative interaction and study its effect on supersymmetry breaking in the presence of a symmetry-breaking potential. $P(X)$-type higher-derivative interactions have been shown to unleash new branches of the theory that are not continuously related to the canonical theory \cite{Koehn:2012ar}, and here we find that the supersymmetric defects behave quite differently depending on which branch of the theory we are on.

\section{Higher-derivative supersymmetric gauge theory}
Since we are interested in Abelian vortices, we begin by constructing the supersymmetric extension of the two-derivative Abelian Higgs model (cf.~\cite{Wess:1992cp}). Employing the notation of Wess and Bagger \cite{Wess:1992cp} in $d=4$, $N=1$ super-Minkowski space $\mathbb{M}^{4|4}$ with signature $(-+++)$,
we consider a single vector superfield $V$ and $m$ chiral superfields $\Phi_i$ with $U(1)$ charges $t_i$, and write the superspace Lagrangian density as
\begin{align}
 \cL_X\!\equiv\frac14\Big(W^\alpha W_\alpha|_{\theta^2}+\bar{W}_{\dot\alpha}\bar{W}^{\dot\alpha}|_{\bar\theta^2}\Big)+\Phid_l\e^{t_lV}\Phi_l|_{\theta^2\bar\theta^2}+\Big[\left(\frac12m_{ij}\Phi_i\Phi_j+\frac13g_{ijk}\Phi_i\Phi_j\Phi_k\right)|_{\theta^2}+H.c.\Big],
 \label{L2}
\end{align}
where repeated indices are to be summed over, and where
\be
W_\alpha\equiv-\frac14\bar{D}^2 D_\alpha V
\ee
is the field strength for the chiral superfield. The first summand in~(\ref{L2}) is the supersymmetric gauge-invariant generalization of the Lagrangian for a free vector field. The vector-superfield multiplet $V$ in the Wess-Zumino (WZ) gauge (cf.~e.g.~\cite[(6.6)]{Wess:1992cp}) reads
\begin{align}
V=-\theta\sigma^m\bar\theta v_m(x)+\i\theta^2\bar\theta\bar\lambda(x)-\i\bar\theta^2\theta\lambda(x)+\frac12\theta^2\bar\theta^2D(x) \ .
\end{align}
Invariance under the $U(1)$ symmetry requires $m_{ij}=0$ if $t_i+t_j\neq0$ and $g_{ijk}=0$ if $t_i+t_j+t_k\neq0$.  All that remains to complete the supersymmetrization of the Abelian Higgs model is to choose a superpotential, which we will do at the end of this section.

We now construct a supersymmetric, gauge-invariant higher-derivative extension of this theory. A general treatment is quite complicated and obscures the essential features. To avoid this, we focus on the extension of the simplest higher-derivative term, $X^2$. Consider the superfield expression
\begin{align}\label{hdgaugesf}
D\G_iD\G_j\bar{D}\G_k{}^\dagger\bar{D}\G_l{}^\dagger\ ,
\end{align}
where we have defined
\be
\G_i\equiv\Phi_i\e^{t_iV} \ .
\ee
The expression~\eqref{hdgaugesf} is gauge invariant because the local $U(1)$ rotation angle is a chiral multiplet, $\bar{D}_{\dot\alpha}\Lambda=0$. With $(tA)_i\equiv t_iA_i$ and the subscript ``$\zf$'' denoting that the fermion fields have been set to zero, the component expansion of \eqref{hdgaugesf} follows from
\begin{align}
\frac1{16} D\G_iD\G_j&\bar{D}\G_k{}^\dagger\bar{D}\G_l{}^\dagger|_{\theta^2\bar\theta^2,\zf}\notag\\
=&\ \p^mA^{}_i\p_mA^{}_j\p^nA^{*}_k\p_nA^*_l-2F^{}_{(i}\p^m A^{}_{j)}F^*_{(k}\p_mA^*_{l)}+F^{}_iF^{}_jF^*_kF^*_l\notag\\
&+\i v_m\big(F_{(i}^{}\p^mA_{j)}^{}F_{(k}^*(tA^*)^{}_{l)}-F_{(i}^{}(tA)^{}_{j)}F_{(k}^*\p^mA_{l)}^*\big)\notag\\
&+\i v_m\big((tA)^{}_{(i}\p^mA_{j)}^{}\p^nA_k^*\p_nA_l^*-\p^nA^{}_i\p_nA^{}_j(tA^*)_{(k}\p^mA_{l)}^*\big)\notag\\
&-\frac12v^mv_m\big((tA)^{}_{(i}F^{}_{j)}(tA^*)^{}_{(k}F^*_{l)}+\frac12(tA)^{}_i(tA)^{}_j\p^nA_k^*\p_nA^*_l+\frac12\p^nA_i^{}\p_nA^{}_j(tA^*)^{}_k(tA^*)^{}_l\big)\notag\\
&+(tA)^{}_{(i}\p^mA^{}_{j)}v_m(tA^*)^{}_{(k}\p^nA^*_{l)}v_n\notag\\
&+\frac{\i}4v^mv_mv^n\big((tA)^{}_i(tA)^{}_j(tA)^*_{(k}\p_nA^*_{l)}-(tA)^{}_{(i}\p_nA^{}_{j)}(tA)^*_k(tA)^*_l\big)\notag\\
&+\frac1{16}(tA)^{}_i(tA)^{}_j(tA^*)^{}_k(tA^*)^{}_l(v^mv_m)^2 \ , \label{compX2}
\end{align}
where the auxiliary field $F$ describes the highest component of the superfield $\Phi$. The component expansion \eqref{compX2} shows that \eqref{hdgaugesf} is a gauge-invariant supersymmetric extension for $X^2$. Let us define
\be
\cL_{X^2}\equiv\frac{\tau}{16}D\G_iD\G_j\bar{D}\G_k{}^\dagger\bar{D}\G_l{}^\dagger T^{}_{ijkl} \ ,
\ee
where $T_{ijkl}$ is symmetric under $i\leftrightarrow j$ and $k\leftrightarrow l$ and shall for the present purposes consist simply of a combination of Kronecker symbols. The full action including both $\cL_X$ and $\cL_{X^2}$, with no Fayet-Iliopoulos term, then takes the form
\begin{align}\label{Lag}
S|_\zf=&\int\d^4x\Big[-|\cD A_i|^2+|F_i|^2+F_iW_{,A_i}+F^*_iW^*_{,A_i^*}-\frac14v^{mn}v_{mn}\notag\\
&\hspace{3.4em}+\tau\Big(\cD^mA^{}_i\cD_mA^{}_j\bcD^nA^{*}_k\bcD_nA^*_l-2F^{}_i\cD^mA^{}_jF^*_k\bcD_mA^*_l+F^{}_iF^{}_jF^*_kF^*_l\Big)T^{}_{ijkl}\Big] \ ,
\end{align}
where $W$ is the holomorphic superpotential and
\begin{align}
\cD_m A^{\phantom{*}}_i\equiv&\ \p_m A^{\phantom{*}}_i+\frac{\i}2v_m(tA^{\phantom{*}})^{}_i\\
\bar{\cD}_m A^*_i\equiv&\ \p_m A^*_i-\frac{\i}2v_m(tA^*)_i \\
v_{mn}\equiv &\ \p_mv_n-\p_nv_m \\ 
(\cD A_i)^2\equiv&\ \eta^{mn}\cD_m A_i\cD_nA_i \\
|\cD A_i|^2\equiv&\ \cD^mA_i^{}\bcD_mA^*_{i} \\
|F_i|^2\equiv&\ F^{}_iF_i^* \ . 
\end{align}
Let us note that the potential in \eqref{Lag} reads
\be
V\equiv\Big[-|F_j|^2-F_jW_{,A_j}-F^*_jW^*_{,A_j^*}-\tau F^{}_iF^{}_jF^*_kF^*_lT^{}_{ijkl}\Big]_\text{pot}\ ,
\ee
where the subscript ``pot'' is appended because the expression on the right-hand side could also contain kinetic terms, which should be omitted.

According to \eqref{Lag}, the equations of motion for the auxiliary fields $F_i$ are given by
\be
F^{}_i+W^*_{,A_i^*}+2\tau F^{}_j\Big(F^{}_kF^*_l-\cD^mA^{}_k\bcD_mA^*_{l}\Big)T^{}_{jkil}=0 \ .
\ee
We choose $T_{ijkl}$ such that the values of all four indices are restricted to be equal to each other, and obtain
\be\label{cube}
F^{}_i+W^*_{,A_i^*}+2\tau F_i\Big(|F^{}_i|^2-|\cD A^{}_i|^2\Big)=0 \ ,
\ee
where in \eqref{cube} and in the remainder of this article, there is no summation on doubly-occurring $i,j,\ldots$ indices. Multiplication of \eqref{cube} with $F^*_i$ reveals
\be
F^*_iW^*_{,A_i^*}=F^{}_iW^{}_{,A_i^{}} \ .
\ee

Finally, in order to break the gauge symmetry, one may either induce spontaneous symmetry breaking (SSB) through an appropriate choice of potential, or one may rely on a non-vanishing Fayet-Iliopoulos term. We choose the first path and construct a model with chiral superfields that each feature higher-derivative interactions. At least three such fields are required to break the gauge symmetry; two charged fields $\Phi_\pm$ with $U(1)$ charges $q_\pm=\pm1$, plus a neutral field $\Phi_0$. We choose the potential (cf.~\cite{Davis:1997bs})
\be\label{pot}
W(\Phi_\pm)=\mu\Phi_0(\Phi_+\Phi_--\eta^2) \ ,
\ee
where $\eta$ is a real dimensionless parameter, and $\mu$ is a real parameter with dimensions of mass. In general, non-vanishing vacuum expectation values (VEVs) of auxiliary fields induce supersymmetry breaking, while non-vanishing VEVs of dynamical scalar fields lead to gauge-symmetry breaking.\footnote{The supersymmetric ghost condensate poses a higher-derivative counterexample: even without the input of a superpotential, the ghost-condensate vacuum spontaneously breaks supersymmetry, and there it is the scalar field that acquires a nonzero vacuum expectation value, leading to the fermion transforming inhomogeneously and thus breaking supersymmetry~\cite{Khoury:2010gb,Koehn:2012te}.} In the presence of a potential, Eqn.~\eqref{cube} can be solved exactly via Cardano's formula.
The resulting expression in terms of cube roots is however too cumbersome to be put to practical use. We therefore approximate the solution for small $\tau$, following the approach in~\cite{Koehn:2012ar}. Because \eqref{cube} is a cubic equation, one obtains three different solution branches, as discussed in \cite{Koehn:2012ar,Koehn:2014zoa}. Selecting the ordinary solution branch
\be
F_i=-W^*_{,A_i^*}+2\tau (W^*_{,A_i^*}){}^2W^{}_{,A_i^{}}-2\tau W^*_{,A_i^*}|\cD A_i|^2+\mac O(\tau^2) \ ,
\ee
equation~\eqref{Lag} may be written to first order in $\tau$ as
\begin{align}\label{smalltauLag}
S|_\zf=\int\d^4x\sum_{i\in\{0,\pm\}}\Big[&-|\cD A_i|^2-|W_{,A_i}|^2-\frac14v^{mn}v_{mn}\notag\\
&+\tau\Big((\cD A^{}_i)^2(\bcD A^{*}_i)^2-2|W_{,A_i}|^2|\cD A^{}_i|^2+|W_{,A_i}|^4\Big)\Big] \ .
\end{align}
One may then proceed to derive the equations of motion for $A_i$ and $v_m$ from \eqref{smalltauLag}.


\section{Cosmic-String Solutions}

One obvious way to construct cosmic-string solutions to the full model is to solve the complete coupled scalar, vector and fermion equations of motion with the appropriate boundary conditions yielding a string background. In practice this is not the most convenient path to take. Instead we follow the procedure of~\cite{Davis:1997bs}, and begin by setting the fermions to zero at first and constructing a cosmic-string solution in the purely bosonic sector of the theory. We will then use supersymmetry transformations to obtain the general fermion solutions in terms of the background string fields.

The cosmic-string ansatz is
\begin{align}
A_0=&\ 0\\
A_+=&\ A_-^*=\eta\e^{\i n\varphi}f(r)\\
v_\mu=&-\frac2{g}n\frac{a(r)}{r}\delta^\varphi_\mu \ ,
\end{align}
where we have included the coupling $g$ to be general, but have from the start set $g=1$.
The equations of motion to first order in $\tau$ reduce for this ansatz to
\begin{align}
f''+\frac{f'}{r}-\frac{n^2}{r^2}f(1-a)^2=&\ \mu^2\eta^2(f^2-1)f-2\tau\mu^4\eta^5(f^2-1)^3f\notag\\
&+2\tau\eta^2\Big(-\frac{n^4}{r^4}f^3(1-a)^4-\frac{n^2}{r^2}ff'^2(1-a)^2+\frac1{r}f'^3+3f'^2f''\notag\\
&\hspace{3.8em}+\frac{n^2}{r^3}f^2f'(1-a)^2-\frac{n^2}{r^2}f^2f''(1-a)^2+2\frac{n^2}{r^2}f^2f'(1-a)a'\Big)\\
a''-\frac{a'}{r}+\frac{a}{r^2}=&-\eta^2f^2(1-a)+2\tau\eta^4f^4\frac{n^2}{r^2}(1-a)^3-2\tau\eta^4f^2f'^2(1-a) \ .
\end{align}
These are second-order equations of motion and can be solved numerically in a standard way for suitable values of the constants, subject to the boundary conditions
\begin{align}
f(0)=a(0)=&\ 0\\
\lim_{r\ra\infty}f(r)=\lim_{r\ra\infty}a(r)=&\ 1 \ .
\end{align}
The chosen string ansatz implies to first order in $\tau$ that
\begin{align}
F_\pm=&\ 0\\
F_0=&-W^*_{,A_0^*}+2\tau W^*_{,A_0^*}{}^2W^{}_{,A^{}_0}=-\mu\eta^2(f(r)^2-1)+2\tau\mu^3\eta^6(f(r)^2-1)^3\ .
\end{align}
Now, assuming we have these solutions in hand, we follow~\cite{Davis:1997bs} and seek the fermionic solutions in terms of the background string fields via the supersymmetry transformations. These transformations are given by $G=\e^{\xi Q+\bar\xi\bar Q}$, with Grassmann parameters $\xi^\alpha$ and supersymmetry algebra generators
\begin{align}
Q_\alpha=&\diffp{\th^\alpha}-\i\s^m_{\a\dot\a}\tb^{\dot\a}\p_m\\
\bar{Q}^{\dot\alpha}=&\diffp{\tb_{\dot\alpha}}-\i\bar\s^{m\dot\a\a}\th_\a\p_m \ .
\end{align}
Left-moving superconducting currents, if they exist, flow along the string at the speed of light and take the form
\be
\Psi_i=\chi_i(r,\varphi)\e^{h(z+t)} \ ,
\ee
with $h$ an arbitrary real function. Our central question is whether such supercurrents exist in the presence of the higher-derivative interactions.
While the supersymmetry transformation on $\lambda$ is not affected by the higher-derivative terms, those on $\psi_i$ are -- because they contain $F$ according to
\begin{align}
\delta_\xi\psi_i=&\i\sqrt2\sigma^m\bar\xi \cD_{m}A_i+\sqrt2\xi F_i \ .
\end{align}
For the case at hand, this means
\begin{align}
(\delta_\xi\psi_0)_\alpha=&\sqrt2\xi_\alpha F_0=-\sqrt2\xi_\alpha\mu\eta^2(f^2-1)+2\tau\sqrt2\xi_\alpha\mu^3\eta^6(f^2-1)^3\label{susy1-ord}\\
(\delta_\xi\psi_\pm)_\alpha=&\sqrt2\Big(\i f'\sigma^r\mp \frac{n}{r}(1-a)f\sigma^\varphi\Big)_{\alpha\dot\alpha}\bar\xi^{\dot\alpha}\eta\e^{\pm\i n\varphi}\label{susy2-ord} \ .
\end{align}
We see that the $k$-defect is not invariant under these transformations, and thus breaks supersymmetry. Because $\tau$ is a small parameter, the situation is qualitatively the same as in \cite{Davis:1997bs}: the conditions $f^2=1$, $f'=0$ and $a=1$ all hold outside of the string core and thus supersymmetry is restored there.

However, for higher-derivative supersymmetric theories, this is not the end of the story. One of the most interesting features of these models is that the cubic equation for the auxiliary field yields different branches of the theories upon replacement of the auxiliary field solution in the Lagrangian \eqref{Lag}. The non-ordinary branches for small $\tau$ are given by
\begin{align}
F_{i,\text{non}}=&\pm\frac{\i}{\sqrt{2\tau}}\sqrt\frac{W^*_{,A_i^*}}{W_{,A_i}}+\frac12W^*_{,A_i^*}\mp\i\sqrt\frac{\tau}2\sqrt\frac{W^*_{,A_i^*}}{W_{,A_i}}|\cD A_i|^2+\mac{O}(\tau) \ .
\end{align}
Upon insertion into the Lagrangian, we obtain
\begin{align}
S|_\zf=&\sum_{i\in\{0,\pm\}}\int\d^4x\Big[-4|\cD A_i|^2+\frac32|W_{,A_i}|^2+\frac9{4\tau}-\frac14v^{mn}v_{mn}+\mac O(\tau)\Big] \ .
\end{align}
As is typical on these new branches, there appears a term in the potential that is inversely proportional to $\tau$, signalling that the new theory is not continuously connected to the ordinary branch for small $\tau$. Neglecting the constant term, the leading-order potential
\be
V=-\frac32\sum_{i\in\{\pm,0\}}|W_{,A_i}|^2
\ee
is not bounded below, and therefore, instead, we proceed to choose the ordinary branch for $F_0$ and non-ordinary branches for $F_\pm$. In this case, the action reads
\begin{align}
S|_\zf=&\sum_{i=\pm}\hspace{-.1em}\int\hspace{-.3em}\d^4x\Big[-|\cD A_0|^2-|W^{}_{,A_0}|^2+\tau (\cD A^{}_0)^2(\bcD A^*_0)^2-2\tau |W^{}_{,A_0}|^2|\cD A_0|^2+\tau|W^{}_{,A_0}|^4-\frac14v^{mn}v_{mn}\notag\\
&\hspace{4.6em}-4|\cD A_i|^2+\frac32|W^{}_{,A_i}|^2+\tau(\cD A^{}_i)^2(\bcD A_i^*)^2+4\tau|\cD A_i^{}|^4\notag\\
&\hspace{4.6em}-\frac{\tau}2|W^{}_{,A_i}|^2|\cD A_i|^2+\frac{\tau}{16}|\cD A_i|^4+\frac3{2\tau}\Big] \ .
\end{align}
As above, we can find fermion solutions in terms of the bosonic background string fields via the supersymmetry transformations. The latter are now given by
\begin{align}
(\delta_\xi\psi_0)_\alpha=&\sqrt2\xi_\alpha F_0=-\sqrt2\xi_\alpha\mu\eta^2(f^2-1)+2\tau\sqrt2\xi_\alpha\mu^3\eta^6(f^2-1)^3\label{susy1-new}\\
(\delta_\xi\psi_\pm)_\alpha=&\sqrt2\Big(\i f'\sigma^r\mp \frac{n}{r}(1-a)f\sigma^\varphi\Big)_{\alpha\dot\alpha}\bar\xi^{\dot\alpha}\eta\e^{\pm\i n\varphi}\pm\frac{\i\e^{\pm\i n\varphi}}{\sqrt\tau}\xi_\alpha\Big(1-\tau\eta^2\big(f'^2+\frac{n^2}{r^2}(1-a)^2f^2\big)\Big)
\label{susy2-new}
\end{align}
We see that, contrary to the case of canonical kinetic term defects~\cite{Davis:1997bs}, an important result is that in general, for $k$-defects, supersymmetry breaking and zero modes seem not to be confined to the string's core because of the new correction term in $\tau$. The physical significance of the new branches remains unclear, and we refer the reader to recent discussions of this topic in~\cite{Gwyn:2014wna,Ciupke:2015msa}.

Note that one may also study the solution for the different branches of $F$ when $\tau$ is large. However, as has been shown in \cite{Koehn:2012ar}, in flat space the potential becomes irrelevant altogether. Allowing for appropriate field-dependent values of $\tau$ could introduce a potential in a new way, but this is beyond the scope of the present work and we leave this possibility open for future studies.

\section{Summary}

We have studied the microphysics of cosmic-string solutions to variants of supersymmetric Abelian Higgs models with certain higher-derivative interactions. The gauge-invariant higher-derivative interaction term that we have introduced implies cubic equations for the auxiliary field $F$, admitting solutions representing different branches of the supersymmetric theory upon reinsertion into the Lagrangian. Because it is difficult to solve the fermionic equations of motion exactly, we have used the supersymmetry transformation to obtain the fermionic zero modes. In the case of the ordinary branch, supersymmetry remains unbroken outside the string's core, but is broken inside of it, and the higher-derivative interactions merely yield correction terms to the superconducting currents found for canonical supersymmetric strings. However, in the case of the non-ordinary branches, the higher-derivative interactions generate entirely new potential terms, and contrary to the behavior on the ordinary branch, supersymmetry no longer remains unbroken outside the string's core. The existence and physics of new branches in supersymmetric higher-derivative theories has been considered in other settings~\cite{Koehn:2012ar,Farakos:2012qu,Gwyn:2014wna,Ciupke:2015msa}, and the new physics they possibly introduce to superconducting defects provides an additional reason for their study. In future work we will focus on the question of whether these branches and their associated phenomena are ultimately relevant to the dynamics in such theories, both in the present setting and more generally.

\acknowledgments
M.K.~would like to thank Jean-Luc Lehners for helpful discussions and acknowledges the support of the US Department of Energy under contract No.~DE-SC0007901. This work was supported in part by US Department of Energy (HEP) Award DE-SC0013528.

\appendix
\allowdisplaybreaks

\section{Complete expression for supersymmetric $P(X)$}

We supplement the results of this article with the full component expansion for the supersymmetric extension of $P(X)$ theories proposed in~\cite{Khoury:2010gb}, where component expansions were evaluated only up to quadratic order in fields other than $\phi$, the real part of the complex scalar field $A=\frac1{\sqrt2}(\phi+\i\xi)$. It was demonstrated that a supersymmetric extension of the action
\be
S_P=\int\d^4xP(X)=\int\d^4x\sum_{n\in\mathbb{N}^*}a_nX^n
\ee
is given by
\begin{align}
S_P^{\rm SUSY}=&\phantom{+}\int\d^4x\d^4\th\Big[K(\Phi,\Phid)+\frac1{16}\sum_{n\geq2}a_nD\Phi D\Phi \Db\Phid\Db\Phid T^{n-2}\Big]\ ,
\label{superP}
\end{align}
where
\be
T\equiv\frac1{32}\{D,\Db\}(\Phi+\Phid)\{D,\Db\}(\Phi+\Phid)\ .
\ee
In this expression, the standard kinetic term $X\equiv-\frac12(\p\phi)^2$ emerges, in its component expansion, as part of $-a(\phi)\p A\p A^*$, where $a(\phi)=K_{,\Phi\Phid}\mid_{\th=\tb=0}$, and $K$ is chosen accordingly. We find that
\begin{align}
T=&-\frac12(\p\phi)^2-\th\chi_{,m}\phi^{,m}-\tb\bar\chi_{,m}\phi^{,m}\nn\\
&+\th^2\big[\frac14(\p\chi)^2-\frac1{\sqrt2}\phi_{,m}F^{,m}\big]+\tb^2\big[\frac14(\p\bar\chi)^2-\frac1{\sqrt2}\phi_{,m}F^{*,m}\big]\nn\\
&-\th\chi^{,m}\tb\bar\chi_{,m}+\th\sigma^m\tb\phi^{,n}\xi_{,mn}\nn\\
&-\frac{\i}2\th^2\tb\bar\sigma^m\chi_{,mn}\phi^{,n}-\frac{\i}2\tb^2\th\sigma^m\bar\chi_{,mn}\phi^{,n}\nn\\
&+\th^2\tb^2\frac14\big[(\xi_{,mn})^2-\phi_{,m}\p^m\Box\phi\big] \ .
\end{align}
We now employ component expansions of powers of $T$, in terms of powers of $X$, yielding the unwieldy expression
\begin{align}\label{Xn}
\frac1{16}\big(D\Phi&D\Phi\Db\Phid\Db\Phid T^n\big)\mid_{\th^2\tb^2}\nn\\
=\Big\{&(\p A)^2(\p A^*)^2-2|F|^2|\p A|^2+|F|^4+\frac{\i}2\big(\chi_{,n}\sigma^n\bar\sigma^m\sigma^l\bar\chi-\chi\sigma^m\bar\sigma^l\sigma^n\bar\chi_{,n}\big)A_{,m}A^*{}_{,l}\nn\\
&+\i\big(\chi\sigma^m\bar\chi^{,n}-\chi^{,m}\sigma^n\bar\chi\big)A_{,m}A^*{}_{,n}+\frac{\i}2\chi\sigma^m\bar\chi(A^*{}_{,m}\Box A-A_{,m}\Box A^*)\nn\\
&+\frac12(F\Box A-F^{,m}A_{,m})\bar\chi^2+\frac12(F^*\Box A^*-F^{*,m}A^*{}_{,m})\chi^2\!+\!2FA_{,m}\bar\chi\bar\sigma^{mn}\bar\chi_{,n}\!+\!2F^*A^*{}_{,m}\chi\sigma^{mn}\chi_{,n}\nn\\
&+\frac32\i|F|^2(\chi_{,m}\sigma^m\bar\chi-\chi\sigma^m\bar\chi_{,m})+\frac{\i}2\chi\sigma^m\bar\chi(FF^*{}_{,m}-F^*F_{,m})\Big\}X^{n}\nn\\
+\Big\{&\frac{\i}4\chi\sigma^m\bar\chi(FA^*{}_{,m}-F^*A_{,m})\big((\xi_{,mn})^2-\phi_{,m}\p^m\Box\phi\big)+\frac{\i}{2\sqrt2}|F|^2\phi^{,n}\big(F^*\chi\sigma^m\bar\chi_{,mn}+F\bar\chi\bar\sigma^m\chi_{,mn}\big)\nn\\
&-\frac12\chi\s^m\bar\chi_{,p}\chi^{,p}\s^n\bar\chi A_{,m}A^*{}_{,n}-\frac12A_{,m}A^*{}_{,n}\chi\s^m\bar\s^p\s^n\bar\chi\phi^{,q}\xi_{,pq}\nn\\
&-\frac12|F|^2\chi\chi^{,m}\bar\chi\bar\chi_{,m}+\frac12|F|^2\phi^{,m}\xi_{,mp}\chi\s^p\bar\chi-\frac{\i}2(FA_{,m}\bar\chi^2-F^*A^*{}_{,m}\chi^2)\big(\frac12\chi^{,n}\s^m\bar\chi_{,n}+\phi_{,n}\xi^{,mn}\big)\nn\\
&-\frac12\big((\p A^*)^2\chi^2-F^2\bar\chi^2\big)\big(\frac14(\p\bar\chi)^2-\frac{\phi_{,m}}{\sqrt2}F^{*,m}\big)-\frac12\big((\p A)^2\bar\chi^2-F^{*2}\chi^2\big)\big(\frac14(\p\chi)^2-\frac{\phi_{,m}}{\sqrt2}F^{,m}\big)\nn\\
&+\frac{\i}{\sqrt2}\phi^{,n}\Big[-A_{,m}(\p A^*)^2\chi\sigma^m\bar\chi_{,n}+A^*{}_{,m}(\p A)^2\chi_{,n}\sigma^m\bar\chi+|F|^2\big(A^*{}_{,m}\chi\sigma^m\bar\chi_{,n}-A_{,m}\chi_{,n}\sigma^m\bar\chi\big)\Big]\nn\\
&+\frac1{\sqrt2}FA^*{}_{,m}A_{,n}\phi^{,p}\bar\chi\bar\sigma^m\sigma^n\bar\chi_{,p}+\frac1{\sqrt2}F^*A_{,m}A^*{}_{,n}\phi^{,p}\chi_{,p}\sigma^n\bar\sigma^m\chi\Big\}nX^{n-1}\nn\\
+\Big\{&-\i\chi\sigma^m\bar\chi(FA^*{}_{,m}-F^*A_{,m})\Big[\frac{\i}4\phi^{,n}\phi^{,p}\chi_{,n}\sigma^m\bar\chi_{,mp}+\frac{\i}4\phi^{,n}\phi^{,p}\bar\chi_{,n}\bar\sigma^m\chi_{,mp}+\frac14\phi_{,n}\phi^{,p}\xi^{,nq}\xi_{,pq}\nn\\
&\hspace{12.2em}-\big(\frac14(\p\chi)^2-\frac1{\sqrt2}\phi_{,n}F^{,n}\big)\big(\frac14(\p\bar\chi)^2-\frac1{\sqrt2}\phi_{,n}F^{*,n}\big)\nn\\
&\hspace{12.2em}-\frac18\chi^{,n}\chi^{,p}\bar\chi_{,n}\bar\chi_{,p}+\frac14\chi^{,n}\sigma^q\bar\chi_{,n}\phi^{,p}\xi_{,pq}\Big]\nn\\
&+\frac1{\sqrt2}|F|^2F^*\phi^{,p}\Big[\chi\chi_{,p}\big(\frac14(\p\bar\chi)^2-\frac1{\sqrt2}\phi_{,n}F^{*,n}\big)+\frac12\chi\chi^{,n}\bar\chi_{,p}\bar\chi_{,n}-\frac12\chi\sigma^m\bar\chi_{,n}\phi^{,n}\xi_{,mp}\Big]\nn\\
&+\frac1{\sqrt2}|F|^2F\phi^{,p}\Big[\bar\chi\bar\chi_{,p}\big(\frac14(\p\chi)^2-\frac1{\sqrt2}\phi_{,n}F^{,n}\big)+\frac12\bar\chi\bar\chi^{,n}\chi_{,p}\chi_{,n}+\frac12\bar\chi\bar\sigma^m\chi_{,n}\phi^{,n}\xi_{,mp}\Big]\nn\\
&+\frac12A_{,m}A^*{}_{,n}\phi^{,p}\phi^{,q}\chi\s^m\bar\chi_{,q}\chi_{,p}\s^n\bar\chi+\frac12|F|^2\phi^{,m}\phi^{,n}\chi\chi_{,m}\bar\chi\bar\chi_{,n}\nn\\
&+\frac{\i}4\phi^{,m}\phi^{,n}\chi_{,m}\s^p\bar\chi_{,n}(FA_{,p}\bar\chi^2-F^*A^*{}_{,p}\chi^2)\nn\\
&+\frac18\big((\p A^*)^2\chi^2-F^2\bar\chi^2\big)\bar\chi_{,n}\bar\chi_{,p}\phi^{,n}\phi^{,p}+\frac18\big((\p A)^2\bar\chi^2-F^{*2}\chi^2\big)\chi_{,n}\chi_{,p}\phi^{,n}\phi^{,p}\Big\}n(n-1)X^{n-2}\nn\\
+\Big\{&-\i\chi\sigma^m\bar\chi(FA^*{}_{,m}-F^*A_{,m})\Big[\chi_{,n}\chi^{,q}\bar\chi_{,p}\bar\chi_{,q}\phi^{,n}\phi^{,p}-\chi_{,n}\sigma^q\bar\chi_{,p}\phi^{,n}\phi^{,p}\phi^{,r}\xi_{,qr}\nn\\
&\hspace{12.2em}+\phi^{,n}\phi^{,p}\chi_{,n}\chi_{,p}\big(\frac14(\p\bar\chi)^2-\frac1{\sqrt2}\phi_{,q}F^{*,q}\big)\nn\\
&\hspace{12.2em}+\phi^{,n}\phi^{,p}\bar\chi_{,n}\bar\chi_{,p}\big(\frac14(\p\chi)^2-\frac1{\sqrt2}\phi_{,q}F^{,q}\big)\Big]\nn\\
&-\frac1{\sqrt2}|F|^2F^*\chi\chi_{,m}\bar\chi_{,n}\bar\chi_{,p}\phi^{,m}\phi^{,n}\phi^{,p}-\frac1{\sqrt2}|F|^2F\bar\chi\bar\chi_{,m}\chi_{,n}\chi_{,p}\phi^{,m}\phi^{,n}\phi^{,p}\Big\}\frac{n!}{4(n-3)!}X^{n-3}\nn\\
+\Big\{&\i\chi\sigma^m\bar\chi(FA^*{}_{,m}-F^*A_{,m})\chi_{,n}\chi_{,p}\bar\chi_{,q}\bar\chi_{,r}\phi^{,n}\phi^{,p}\phi^{,q}\phi^{,r}\Big\}\frac{n!}{16(n-4)!}X^{n-4} \ .
\end{align}
The combination of this expression with the well-known supersymmetric extension for $X$ (e.g.~\cite{Wess:1992cp}) gives the complete supersymmetric $P(X)$ with superpotential $W$ as
\begin{align}\label{susyPfull}
S_P^{\rm SUSY}=&\int\d^4x\d^2\th \big[-\frac18\Db^2K(\Phi,\Phid)+W(\Phi)\big]+H.c.+\frac1{16}\sum_{n\geq2}a_n\int\d^4x\d^2\th\d^2\tb D\Phi D\Phi \Db\Phid\Db\Phid T^{n-2}\nn\\
=&\int\d^4x\Big[-K_{,AA^*}(\p A)(\p A^*)+K_{,AA^*}|F|^2-\frac{\i}2K_{,AA^*}\bar\chi\bar\s^m\chi_{,m}+\frac{\i}2K_{,AA^*}\bar\chi_{,m}\bar\s^m\chi\nn\\
&\hspace{3.3em}-\frac12FK_{,AA^*A^*}\bar\chi^2-\frac12F^*K_{,AAA^*}\chi^2+\frac14K_{,AA^*AA^*}\chi^2\bar\chi^2\nn\\
&\hspace{3.3em}+FW_{,A}+F^*W^*_{,A^*}-\frac12W_{,AA}\chi^2-\frac12W_{,A^*A^*}\bar\chi^2\nn\\
&\hspace{3.3em}+\frac1{16}\sum_{n\geq2}a_n\int\d^2\th\d^2\tb D\Phi D\Phi\Db\Phid\Db\Phid T^{n-2}\Big] \ ,
\end{align}
where we abbreviate e.g.~$K_{,AA^*}\equiv K_{,\Phi\Phid}|_{\th=\tb=0}$. Note that the above expressions can be straightforwardly generalized to the case with more than one scalar superfield~\cite{Koehn:2012ar}.

\bibliographystyle{apsrev-title}
\bibliography{supersymmetric-k-defects}

\end{document}